\documentclass[11pt,a4paper]{article}
\usepackage{amsmath,amssymb,mathtools,amsthm}
\usepackage{authblk}
\usepackage{graphicx}

\usepackage{todonotes}
\usepackage{subfigure}

\usepackage{tikz}
\usepackage[T1]{fontenc}
\usepackage{color}
\usepackage{xhfill}

\usetikzlibrary{calc}
\usetikzlibrary{shapes.geometric, arrows,matrix}
\usetikzlibrary{decorations.markings}
\tikzset{nodeStyle/.style = {circle,draw,minimum size=30pt}}
\tikzset{arrowStyle/.style = {-latex}}

\newcommand{\ket}[1]{\ensuremath{|#1\rangle}}
\newcommand{\bra}[1]{\ensuremath{\langle#1|}}
\newcommand{\ketbra}[2]{\ensuremath{\ket{#1}\bra{#2}}}
\newcommand{\proj}[1]{\ensuremath{\ket{#1}\bra{#1}}}

\newcommand{\LL}{\mathcal{L}}
\newcommand{\MM}{\mathcal{M}}

\newcommand{\1}{{\rm 1\hspace{-0.9mm}l}}
\newcommand{\Id}{\1}
\newcommand{\ii}{\mathrm{i}}
\newcommand{\dd}{\mathrm{d}}

\newcommand{\kron}{\otimes}
\newcommand{\con}[1]{\bar{#1}}
\newcommand{\vecc}[1]{ | #1 \rangle \rangle}

\newcommand{\ie}{\emph{i.e.\/}}

\newcommand{\indeg}{\operatorname{indeg}}

\newtheorem{lemma}{Lemma}
\newtheorem{theorem}{Theorem}

\title{Superdiffusive quantum stochastic walk definable of arbitrary directed graph}
\author[1]{Krzysztof Domino}
\author[1,2]{Adam Glos\thanks{aglos@iitis.pl}}
\author[1,2]{Mateusz Ostaszewski}
\affil[1]{Institute of Theoretical and Applied Informatics,\protect\\ Polish Academy of 
Sciences, \protect\\ Ba{\l}tycka 5, 44-100 Gliwice, Poland}

\affil[2]{Silesian University of Technology, \protect\\  Institute of Informatics,\protect\\
Akademicka 16, 44-100 Gliwice, Poland}

\date{}
\begin{document}
	\maketitle

	\begin{abstract}
		In this paper we define a quantum stochastic walk on arbitrary
		directed graph with super-diffusive propagation on a line graph. Our model is
		based on global environment interaction QSW, which is known to have ballistic
		propagation. However we discovered, that in this case additional amplitude
		transitions occur, hence graph topology is changed into moral graph. Because of
		that we call the effect a spontaneous moralization. We propose a general
		correction scheme, which is proved to remove unnecessary transition and thus to
		preserve the graph topology. In the end we numerically show, that
		super-diffusive propagation is preserved. Because of that our new model may be
		applied as effective evolution on arbitrary directed graph.
	\end{abstract}

	\paragraph{keywords}Quantum stochastic walk, superdiffusive propagation, spontaneous moralization,
		directed graphs

	\section{Introduction} 
	
	Quantum stochastic walk (QSW) is a model of continuous quantum walk, which
	generalizes both random walk and quantum walk \cite{whitfield2010quantum}. QSW
	is governed by the Gorini-Kossakowski-Sudarshan-Lindblad (GKSL) master equation
	\cite{kossakowski1972quantum, gorini1976completely, lindblad1976generators},
	which is a general description of a continuous quantum evolution of an open
	system. The motivation to investigate quantum stochastic walk comes from the
	fact that it models a quantum propagation with an environmental interaction
	\cite{whitfield2010quantum}. For this reason, quantum stochastic walks have a
	potential for modelling real-world quantum systems and quantum computers.
	Moreover, there are currently known applications of quantum stochastic walks
	\cite{falloon2016qswalk} such as: the interpolation between quantum and
	classical dynamics for the one-dimensional tight-binding model in the electron
	transport theory \cite{datta1997electronic}, the pure de-phasing scattering
	process \cite{kendon2007decoherence}, the photosynthetic light-harvesting
	\cite{mohseni2008environment, blankenship2013molecular}, and the quantum
	Page-Rank algorithm \cite{sanchez2012quantum, loke2015comparing}.
	
	Generally the advantage of the quantum walk comes from the ballistic propagation
	regime. The significance of quantum walks for quantum algorithms is reviewed in
	\cite{kendon2006random,kempe2003quantum}. However, in the case of quantum
	stochastic walk, the propagation speed depends on the choice of Lindblad
	operators collection. In \cite{bringuier2016central} authors have shown that in
	the case of a local environment interaction where each Lindblad operator
	corresponds to a single edge, the ballistic propagation is broken due to
	decoherence and we observe linear propagation typical for classical random walk.
	However, in the case of a global environment interaction where adjacency matrix
	is the only one Lindblad operator, we have proven analytically
	\cite{domino2016properties} that ballistic propagation property is preserved,
	henceforth we find this case particularly interesting.

	It is natural for modelling quantum walks to `divide' the space into orthogonal
	subspaces and combine them with the vertices of the graph. The edge set is
	enclosed in the evolution operator construction. For example in the case of
	continuous quantum walk each canonical state corresponds to different vertex and
	Hamiltonian chosen is usually laplacian or adjacency matrix of the graph. Thanks
	to such assumption we bound the evolution to the graph structure.
	
	Similar attempt has been chosen for quantum stochastic walks 
	\cite{whitfield2010quantum}. However in the
	case of the global environment interaction there appears additional transitions between some
	disconnected vertices, and hence the graph topology is changed. It can lead to
	unintuitive results. As an example suppose we start in the state corresponding
	to some vertex $v$ and there is no path from it to other vertex~$w$. There may
	be nonzero probability of measuring the vertex $w$ after evolution. This 
	phenomenon would disturb some quantum algorithms, if we simply replace local 
	Lindblad operators by a global one to exploit a speed-up due to a ballistic 
	propagation. Moreover the graph topology change has impact on ranking 
	algorithms, since adding or removing edges changes degree of the vertices and 
	hence may change the overall result. Therefore it is
	important to determine how the graph is changed in the quantum
	stochastic walk evolution. In the paper we have discovered, that the additional
	transition is possible, whether vertices have common child. Hence the graph
	changes into its moral graph, \ie{} a graph where all parents with a common 
	child are connected \cite{jensen1996introduction}. 
	Because of that we call this effect the
	spontaneous moralization. In addition we propose a new model of quantum
	stochastic walk, which follows precisely the graph structure.
	
	It is beneficial to determine, if the fast propagation of the walk comes from
	the additional links allowance. To do so, we considered an evolution on
	undirected line graph with our new model and recomputed the result from
	\cite{domino2016properties}. The main result of this work is a fact that our
	model maintains fast propagation and hence can be used for the quantum search
	algorithm on directed graphs. Such propagation appears to be ballistic, at least
	for some intensities of interaction.
	
	The paper is organized as follows. In Sec.~\ref{sec:preliminaries} we provide
	preliminaries and describe the research problem. In Sec.~\ref{sec:new-model} we
	propose a new model of quantum stochastic walk, in which the spontaneous
	moralization does not occur. In Sec.~\ref{sec:hurst} we provide evidence that
	our model maintains fast propagation. In Sec.~\ref{sec:conclusion} we conclude
	our results.

	\section{Preliminaries} \label{sec:preliminaries}
	
	\subsection{GKSL master equation and quantum stochastic walks}  
	
	To define quantum stochastic walks in general, let us start with the
	Gorini-Kossakowski-Sudarshan-Lindblad (GKSL) master equation, that is a
	differential equation of the form \cite{kossakowski1972quantum,
		lindblad1976generators, gorini1976completely}
	\begin{equation}
	\frac{\textrm{d}}{\textrm{d}t}\varrho =\MM (\varrho)= -\ii [H, \varrho] + 
	\sum_{L\in 
		\LL} \left(L \varrho 
	L^\dagger - \frac12 \{L^\dagger L, \varrho\} \right), 
	\label{eq:GKSL-master-equation}
	\end{equation}
	where $\{A, B\}$ is the anticommutator and $\MM$ is the evolution superoperator.
	Here $H$ is the Hamiltonian, which describes the evolution of the closed system,
	and $\LL$ is the collection of Lindblad operators, which describes the evolution
	of the open system. This master equation describes general continuous evolution
	of mixed quantum states.
	
	In the case where $H$ and $\LL$ do not depend on time, we say that Eq.
	\eqref{eq:GKSL-master-equation} describes the Markovian evolution of the system.
	Henceforth, we can solve the differential equation analytically: if we choose
	initial state $\varrho_0$, 
	\begin{equation}
	\vecc{\varrho_t} = S_t \vecc{\varrho_0}, \label{eq:evolution}
	\end{equation}
	where
	\begin{equation}
	S_t = \exp\left[-\ii t\left(H \kron \Id - \Id \kron \con H \right) +
	t\sum_{L\in\LL} \left ( L \kron \con L - \frac{1}{2} L^\dagger L \kron \Id - \Id
	\kron L^\top \bar L \right )\right],
	\end{equation}
	and $\vecc{\,\cdot\,}$ denotes the vectorization of the matrix (see eg.
	\cite{miszczak2011singular}). Using Eq.~\eqref{eq:GKSL-master-equation} we can
	define the propagation from $\ketbra{v}{w}$ to $\ketbra{v'}{w'}$ as
	\cite{whitfield2010quantum}
	\begin{equation}\label{eq:prop}
	\begin{split}
	\MM_{vw} ^{v'w'} &= \delta_{ww'}\bra{v'}\left(-\ii H 
	-\frac{1}{2}\sum_{L\in\LL}L^\dagger L\right) 
	\ket{v}+\delta_{vv'}\bra{w}\left(\ii H 
	-\frac{1}{2}\sum_{L\in\LL}L^\dagger L\right) \ket{w'} 
	+\\
	&\phantom{=\ }+\sum_{L\in\LL}\bra{v'}L\ket {v} \bra {w} L^\dagger\ket {w'},
	\end{split}
	\end{equation}
	where $\delta_{ij}$ is the Kronecker delta.
	
	The GKSL master equation was used for defining quantum stochastic walk. It is
	generalization of both classical random walk and quantum walk
	\cite{whitfield2010quantum}. Both $H$ and $\LL$ correspond to the graph
	structure, however one may verify that at least a choice of Lindblad
	operators may be non-unique \cite{whitfield2010quantum}. Suppose we have the
	undirected graph $G=(V,E)$. Two main models can be distinguished. In both models
	we choose the Hamiltonian $H$ to be the adjacency matrix of the graph. In the
	\emph{local environment interaction case} each Lindblad operator corresponds to
	a single edge, $ \LL = \{\ketbra{w}{v}\colon \{v,w\}\in E\}, $ and in the
	\emph{global environment interaction case} we choose a single Lindblad operator
	$\LL=\{H\}$. In \cite{bringuier2016central} authors have shown that the local
	case leads to the classical propagation of the walk. Oppositely, in the case of
	the global interaction, the ballistic propagation is obtained
	\cite{domino2016properties}. Henceforth, we find the choice of the global
	interaction case particularly interesting and only this case will be analysed.

	\subsection{Research problem and motivation}\label{sec:problem} 
\begin{figure}
		\centering
		\subfigure[\label{fig:intuitive-graph}]{\begin{minipage}[b]{.35\linewidth}
				\centering
				\begin{tikzpicture}[node distance=2.5cm,thick]
				\tikzset{nodeStyle/.style = {circle,draw,minimum size=2.5em}}
				
				\node[nodeStyle] (A)  {$v_1$};
				\node[nodeStyle] (C) [below right of=A] {$v_3$};
				\node[nodeStyle] (B) [above right of=C] {$v_2$};
				
				\tikzset{EdgeStyle/.style   = {thick,-triangle 45}}
				\draw[EdgeStyle] (A) to (C);
				\draw[EdgeStyle] (B) to (C);
				
				\end{tikzpicture}
				
		\end{minipage}}
		\hspace{1cm}
		\subfigure[\label{fig:true-graph}]{\begin{minipage}[b]{.35\linewidth}
				\centering
				\begin{tikzpicture}[node distance=2.5cm,thick]
				\tikzset{nodeStyle/.style = {circle,draw,minimum size=2.5em}}
				
				\node[nodeStyle] (A)  {$v_1$};
				\node[nodeStyle] (C) [below right of=A] {$v_3$};
				\node[nodeStyle] (B) [above right of=C] {$v_2$};
				
				\tikzset{EdgeStyle/.style   = {thick,-triangle 45}}
				\draw[EdgeStyle] (A) to (C);
				\draw[EdgeStyle] (B) to (C);
				
				\tikzset{additionalEdgeStyle/.style = {dashed}}
				\draw[additionalEdgeStyle] (A) -- (B);
				\end{tikzpicture}

		\end{minipage}}
		
		\caption{Visualisation of a directed graph \subref{fig:intuitive-graph} and
			its moral graph \subref{fig:true-graph}.
		}\label{fig:lindbladWalkGraphExample}
	\end{figure}
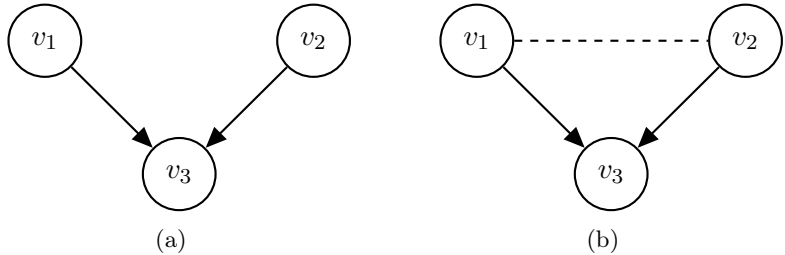

To provide the
	motivation let us consider the following example. Suppose we choose the GKSL
	master equation which consists of the Lindblad operators part only and let us
	consider the graph presented in Fig.~\ref{fig:intuitive-graph}. We choose the
	natural Lindblad operator describing the evolution on this graph as
	\begin{equation}
	L =  \ketbra{v_3}{v_1}+\ketbra{v_3}{v_2}, 
	\label{eq:intuitive-lindblad}
	\end{equation}
	and the initial state $\varrho_0=\ketbra{v_1}{v_1}$. If the Lindblad operator
	describes the evolution on the graph presented in Fig.
	\ref{fig:intuitive-graph}, we should expect $\bra {v_2} \varrho_t\ket{v_2}=0$
	for arbitrary $t\geq0$. However, calculations shows that
	\begin{equation}
	\begin{split}
	\varrho_t &=  \frac{1}{4} e^{-2 t} \left(1+e^t\right)^2 \ketbra{v_1}{v_1}
	+\frac{1}{4} \left(-1+e^{-2 
		t}\right) \ketbra{v_1}{v_2} \\
	&\phantom{\ =}+\frac{1}{4} \left(-1+e^{-2 t}\right) \ketbra{v_2}{v_1} + 
	e^{-t} 
	\sinh ^2\left(\frac{t}{2}\right) \ketbra{v_2}{v_2} \\
	&\phantom{\ =}+  e^{-t} \sinh (t) \ketbra{v_3}{v_3}.
	\end{split}
	\end{equation}
	In a time limit we obtain
	\begin{equation}
	\lim\limits_{t\to\infty}\varrho_t = \frac{1}{4}(\ketbra{v_1}{v_1}  - 
	\ketbra{v_1}{v_2} - \ketbra{v_2}{v_1} +\ketbra{v_2}{v_2}) + \frac{1}{2} 
	\ketbra{v_3}{v_3},
	\end{equation}
	hence we have non-zero probability of measuring the state in node 2. The
	analysis of Eq.~(\ref{eq:prop}), which is demonstrated in detail in
	Sec.~\ref{sec:correct-approach}, shows that an operator from
	Eq.~\eqref{eq:intuitive-lindblad} corresponds to a graph with the new link
	created between $v_1$ and $v_2$, see Fig. \ref{fig:true-graph}. It is because of
	the common child $v_3$, which allows additional propagation inconsistent with
	the original graph structure. For an arbitrary graph, an additional link appears
	between every two disconnected vertices, if they have at least one common child.
	Since such obtained graph is a moral graph of the original one, we call this
	effect the \emph{spontaneous moralization} of a graph. This phenomenon makes 
	the simple use of a global interaction Lindblad operator to a 
	ranking algorithms, such as Page-Rank, impossible.
	
	One should note, that the spontaneuous moralization do not specificaly depends
	on the graph chosen, but on the Lindblad operators. For example if we choose
	local Lindblad operators, we would not observe any additional propagation.
	However, as was mentioned before, local Lindblad are known of slow propagation
	\cite{bringuier2016central}. We are interested in finding such a continuous
	model, in which the effect does not occur and which preserves at least super-diffusive
	propagation. This is done in Sec~\ref{sec:new-model}.

	\section{Quantum stochastic walk on authentic graphs}\label{sec:new-model}
	
	In Sec.~\ref{sec:problem} we considered the spontaneous moralization, caused by
	the Lindblad part of the GKSL equation. In this section we propose new model
	which preserves the original graph structure, and discuss it using only the
	Lindblad part of the GKSL equation. We start with analysis of the GKSL master
	equation in order to determine the causes of the spontaneous moralization in
	Sec.~\ref{sec:spontaneous-moralization}. Then in
	Sec.~\ref{sec:correct-approach}--\ref{sec:symmetry} we propose our model which
	preserve the graphs topology. In Sec.~\ref{sec:correct-approach} we suppress the
	propagation between parents against the graph structure. Next, in Sec.
	\ref{sec:premature-localization} we remove the premature localization effect,
	which occurs in our model. Finally, in Sec. \ref{sec:symmetry} we correct
	asymmetry of frequency distribution of stochastic walks on symmetric graphs.
	
	\subsection{Spontaneous moralization}\label{sec:spontaneous-moralization}
	
	Let us recall the transition rate formula of Lindblad operator with the global
	environment interaction
	\begin{equation}
	\begin{split}
	\MM_{vw} ^{v'w'} &= -\frac12 \delta_{ww'} \bra{v'}L^\dagger L 
	\ket{v}-\frac12 \delta_{vv'}
	\bra{w}L^\dagger L \ket{w'} 
	+\bra{v'}L\ket {v} \bra {w} L^\dagger\ket {w'}.
	\end{split}
	\end{equation}
	One can notice that the only term responsible for the spontaneous moralization
	is $L^\dagger L$. Kronecker deltas cause that one of those terms is non-zero if:
	\begin{enumerate}
		\item $w=w'$, where we have propagation $\MM_{vw}^{v'w}$,
		\item $v=v'$, where we have propagation $\MM_{vw}^{vw'}$,
		\item both $v=v'$ and $w=w'$, which results in $\MM_{vw}^{vw}$.
	\end{enumerate}
	The last case describes the amplitude preservation on the off-diagonal element
	$\ketbra{v}{w}$ of the density matrix. Such preservation of the amplitude is
	consistent with structure of the graph. Moreover, first two cases are
	equivalent, hence we focus on the first case only. This explanation shows, that
	the interaction between not connected vertices occurs only if they have common
	child.
	
	Before we do such we provide another explanation of the spontaneous
	moralization. Let us analyse again the example provided in
	Sec.~\ref{sec:problem}. The Lindblad operator takes the form
	\begin{equation}
	L = \ket{v_3}(\bra{v_1}+\bra{v_2})
	\end{equation}
	and the pure initial state $\ket{v_1}$ can be written as 
	\begin{equation}
	\ket{v_1} = \frac{1}{2} (\ket{v_1}+\ket{v_2}) +\frac{1}{2} 
	(\ket{v_1}-\ket{v_2}).
	\end{equation}
	Note that those two vectors in RHS are orthogonal. Due to decoherent behaviour 
	of
	the Lindblad operator the left part is projected onto $\ket{v_3}$ state, while
	the right vector is preserved\footnote{We would like to thank anonymous reviewer
		for proposing such explanation of the phenomenon.}. Here we can see, that in 
	order
	to propose new model which perseveres the graph topology we need to project all
	vector from the joint subspace of parents subspaces of fixed vertex.
	
	It is worth noting, that such problem does not occur in the local interaction
	case, which is currently deeply investigated
	\cite{falloon2016qswalk,sanchez2012quantum}, however, as it was mentioned
	previously, in this case the model is known to have linear propagation
	\cite{bringuier2016central}. Hence in our opinion it would be beneficial to
	provide a continuous model which can be applied for arbitrary directed graph, 
	and
	possesses super-diffusive propagation.
	
	\subsection{New model} \label{sec:correct-approach}
	
	Suppose we have a directed connected graph $G=(V,E)$ where
	$V=\{v_0,\dots,v_{n-1}\}$. We construct a new graph $\tilde G$ which is
	homomorphic to $G$, \ie{} for which there exists function from $\tilde V$ to $V$
	that maps adjacent vertices of $\tilde G$ to adjacent vertices of $G$. For the
	notation consistency we will now assume, that any part which refers to the graph
	$\tilde G$ will by highlighted by $\sim$.
	
	The graph $\tilde G$ consists of vertex set
	\begin{equation}
	\tilde V = \bigcup_{i=0}^{n-1} [ v_i],
	\end{equation}
	where $[ v_i]=\{\tilde v_i^0 ,\tilde v_i^1,\dots,  
	\tilde v_i^{\indeg(v_i)-1}\}$ iff $\indeg 
	(v_i)>0$ 
	and $[ v_i]=\{\tilde v_i^0\}$ otherwise, and edge set
	\begin{equation}
	(\tilde v_{i}^{k}, \tilde v_{j}^{l})\in \tilde E \iff (v_{i},v_{j})\in E.
	\end{equation} 
	Here $\indeg (v_i)$ is a number of arcs incoming to a vertex $v_i$. %
	In other words we create additional copies of vertices and make a connection
	between them iff their representatives from original graph $G$ are connected.
	
	If we choose $\tilde L$ to be simply an adjacency matrix of $\tilde G$, again we
	may have additional coherence between disconnected nodes. We need to choose
	such entries of the Lindblad operator $\tilde L$ such that $\tilde v_i^l$-th and
	$\tilde v_j^k$-th column are orthogonal, if  $i\neq j$. We choose Lindblad 
	operator $\tilde L$ of the form
	\begin{equation}
	\bra{\tilde v_i^k}\tilde L\ket{\tilde v_j^l} = 
	\begin{cases}\bra k A_{i} \ket j, & (v_j,v_i)\in E,\\
	0, & {\rm otherwise},
	\end{cases} \label{eq:new-lindbladian-entries}
	\end{equation}
	where $A_{i}$ is square matrix of the size $\indeg (v_i)$, such that columns
	form an orthogonal basis. Note, that such choice of $A$ matrices guarantees that
	every vector from the `parents subspace' will be projected onto some child. The
	evolution takes the form
	\begin{equation}
	\frac{\textrm{d}}{\textrm{d}t}\tilde \varrho_t =\tilde L \tilde \varrho \tilde
	L^\dagger - \frac12 \{\tilde L^\dagger \tilde L, \tilde \varrho\}.
	\label{eq:repaired_M}
	\end{equation}
	We define the probability of measuring the state in $v_i$ after time $t$ as
	\begin{equation}
	p(v_i,t) = \sum_{\tilde v\in[v_i]} \bra{\tilde v}\tilde \varrho_t\ket{\tilde 
		v}.
	\end{equation}
	
	\begin{lemma}
		\label{lem:no_moral} Let us take Lindblad operator defined in Eq.
		(\ref{eq:new-lindbladian-entries}). Suppose $\tilde v_i^k,\tilde v_j^l\in
		\tilde V$ are arbitrary vertices such that $i\neq j$. Then $\bra{\tilde
			v_i^k}\tilde L^\dagger \tilde L\ket{\tilde v_j^l}=0$.
	\end{lemma}
	\proof{
		We have
		\begin{equation}{\label{eq::mor}}
		\begin{split}
		\bra{\tilde v_i^k}\tilde L^\dagger \tilde L\ket{v_j^l} &= \sum_{\tilde v\in 
			\tilde V} 
		\bra{\tilde v_i^k}\tilde L^\dagger  \ket {\tilde v}\bra{\tilde v}\tilde 
		L\ket{\tilde v_j^l}\\
		&=\sum_{p=0}^{n-1}\ \sum_{r=0}^{\indeg(v_p)-1}\overline {\bra{\tilde 
				v_p^r}\tilde 
			L\ket 
			{\tilde v_i^k}} \bra{\tilde v_p^r}\tilde L\ket{\tilde v_j^l}\\
		&= \sum_{\substack{p\in\{0,\dots,n-1\}\\(v_j,v_p), (v_i,v_p)\in E}} 
		\sum_{r=0}^{\indeg(v_p)-1} \overline 
		{\bra r A_{p}\ket i} \bra r A_{p} \ket j.
		\end{split}
		\end{equation}
		Columns $A_p\ket{i}$ and $A_p\ket{j}$ are orthogonal which confirms the claim.
	}
	
	\begin{theorem}\label{t:enlarg}
		Let us take two different vertices $\tilde v,\tilde v'$ which has different
		representatives, such that the representatives are not
		connected. Then $\tilde\MM_{\tilde v\tilde w}^{\tilde v'\tilde w}=\tilde
		\MM_{\tilde w\tilde v}^{\tilde w\tilde v'}=0$ for arbitrary $\tilde w$.
	\end{theorem}
	\proof{
		We have 
		\begin{equation}
		\begin{split}
		\tilde \MM_{\tilde v\tilde w}^{\tilde v'\tilde w}&=-\frac{1}{2}\bra{\tilde 
			v'}\tilde L^\dagger\tilde L 
		\ket{\tilde v} 
		+\bra{\tilde v'}\tilde L\ket {\tilde v} \bra {\tilde w} \tilde L^\dagger\ket 
		{\tilde 
			w} \\&= 
		-\frac{1}{2}\bra{\tilde v'}
		\tilde L^\dagger\tilde L 
		\ket{\tilde  v} = 0,
		\end{split}
		\end{equation}
		where the second equality follows from the fact that representatives of $\tilde 
		v$ and $\tilde v'$
		are not connected in original graph $G$, and the last equality follows from
		Lemma~\ref{lem:no_moral}. Proof for $\tilde \MM_{\tilde w\tilde v}^{\tilde
			w\tilde v'}=0$ is analogical.
	}

	From Theorem \ref{t:enlarg}, we conclude that the model defined by
	Eq.~\eqref{eq:repaired_M} evolves on the original graph. Throughout this paper
	we choose $A_{i}$ to be a Fourier matrix, \ie{} $\bra{k}A_i \ket l = \exp \left(
	\frac{2\pi\ii kl}{\indeg (v_i)}\right)$.
	
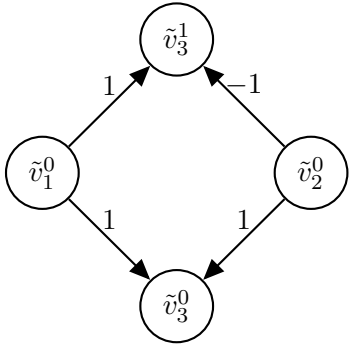
\begin{figure}
		\centering
		
		\begin{tikzpicture}[node distance=2.5cm,thick]
		\tikzset{nodeStyle/.style = {circle,draw,minimum size=2.5em}}
		
		\node[nodeStyle] (A)  {$\tilde v_1^0$};
		\node[nodeStyle] (C0) [below right of=A] {$\tilde v_3^0$};
		\node[nodeStyle] (C1) [above right of=A] {$\tilde v_3^1$};
		\node[nodeStyle] (B) [above right of=C] {$\tilde v_2^0$};
		
		\tikzset{EdgeStyle/.style   = {thick,-triangle 45}}
		\draw[EdgeStyle] (A) to node[above] {1} (C0);
		\draw[EdgeStyle] (B) to node[above] {1}  (C0);
		\draw[EdgeStyle] (A) to node[above] {1} (C1);
		\draw[EdgeStyle] (B) to node[above] {$-1$} (C1);
		
		\end{tikzpicture}\vspace{1em}
		\caption{Visualization of the Lindblad operator in new model. The original
			graph is presented in
			Fig.~\ref{fig:intuitive-graph}.}\label{fig:corrected-graph}
	\end{figure}

	Let us recall the example given in Fig.~\ref{fig:intuitive-graph}. In our model
	the Lindblad operator will be of the form
	\begin{equation}
	\tilde L = \ketbra{\tilde v_3^0}{\tilde v_1^0}+\ketbra{\tilde v_3^1}{\tilde
		v_1^0}-\ketbra{\tilde v_3^1}{\tilde v_2^0}+ \ketbra{\tilde v_3^1}{\tilde v_2^0}.
	\end{equation}
	The operator is presented in Fig.~\ref{fig:corrected-graph}. One can verify
	that, if $\tilde \varrho_0=\proj{\tilde v_1^0}$, then the evolution described by
	Eq.~(\ref{eq:repaired_M}) yields
	\begin{equation}
	\tilde \varrho_t = e^{-2t} \proj{\tilde v_1^0} + \frac{1}{2} e^{-2 t}
	\left(-1+e^{2 t}\right)  (\ket{\tilde v_3^0}+\ket{\tilde v_3^1})(\bra{\tilde
		v_3^0}+\bra{\tilde v_3^1}).
	\end{equation}
	The probability of measuring the state in $v_2$ for arbitrary $t\geq 0$ equals
	0.

	\subsection{Premature localization}\label{sec:premature-localization} 
	
\begin{figure}
		\centering
		\subfigure[\label{fig:premature-localization-example-original}]{
			\begin{minipage}[b]{.45\linewidth}
				\centering
				\begin{tikzpicture}[node distance=2cm,thick]
				\tikzset{nodeStyle/.style = {circle,draw,minimum size=2.5em}}
				
				\node[nodeStyle] (A)  {$v_1$};
				\node[nodeStyle] (C) [above right of=A] {$v_3$};
				\node[nodeStyle] (D) [below right of=A] {$v_4$};
				\node[nodeStyle] (B) [above right of=D] {$v_2$};

				\tikzset{EdgeStyle/.style   = {thick,-triangle 45}}
				\draw[EdgeStyle] (A) to (D);
				\draw[thick] (A) to (C);
				\draw[thick] (A) to (B);
				\draw[thick] (B) to (C);
				\end{tikzpicture}
				
		\end{minipage}}
		\hspace{.5cm}
		\subfigure[\label{fig:premature-localization-example-increased}]
		{\begin{minipage}[b]{.45\linewidth}
				\centering
				\begin{tikzpicture}[node distance=2cm,thick]
				\tikzset{nodeStyle/.style = {circle,draw,minimum size=2.5em}}
				
				\node[nodeStyle] (A0)  {$\tilde v_1^0$};
				\node[nodeStyle] (A1) [above of=A0]  {$\tilde v_1^1$};
				\node[nodeStyle] (C0) [above right of=A1] {$\tilde v_3^0$};
				\node[nodeStyle] (C1) [right of=C0] {$\tilde v_3^1$};
				\node[nodeStyle] (D) [below right of=A0] {$\tilde v_4^0$};
				
				\node[nodeStyle] (B1) [below right of=C1] {$\tilde v_2^1$};
				\node[nodeStyle] (B0) [below of=B1] {$\tilde v_2^0$};

				\tikzset{EdgeStyle/.style   = {thick,-triangle 45}};
				
				\draw[thick] (A0) to (B0);
				\draw[thick] (A0) to (C0);
				\draw[EdgeStyle] (A0) to (D);
				\draw[thick] (A0) to (B1);
				\draw[thick] (A0) to (C1);
				\draw[thick] (B0) to (C0);
				\draw[thick] (B0) to (A1);
				\draw[thick] (B0) to (C1);
				\draw[thick] (C0) to (A1);
				\draw[thick] (C0) to (B1);
				\draw[EdgeStyle] (A1) to (D);
				\draw[thick] (A1) to (B1);
				\draw[thick] (A1) to (C1);
				\draw[thick] (B1) to (C1);
				
				\end{tikzpicture}

			\end{minipage}
		} \caption{The original graph with the premature localization
			\subref{fig:premature-localization-example-original}, and  the graph $\tilde
			G$ based on the original graph
			\subref{fig:premature-localization-example-increased}. Starting from
			$\ketbra{\tilde v_1^0}{\tilde v_1^0}$ we obtain a stationary state which is
			not fully localised in the vertex $\tilde
			v_4^0$.}\label{fig:premature-localization-example}
	\end{figure}
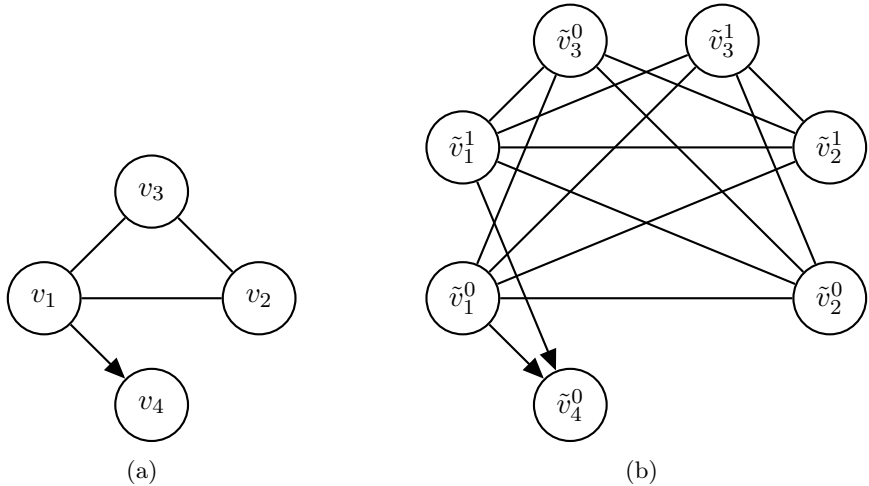

	Through numerical analysis of new model we noticed undesired phenomenon:
	premature localization. For example, let us consider the graph presented in
	Fig.~\ref{fig:premature-localization-example-original}. In our model the
	Lindblad operator will represent the graph presented in
	Fig.~\ref{fig:premature-localization-example-increased}. Its Lindblad operator
	$\tilde L$ has the form
	\begin{equation}
	\tilde L = \begin{bmatrix}
	0 & 1 & 1 & 0 & 0 & 1 & 1\\
	1 & 0 & 1 & 0 & 1 & 0 & 1\\
	1 & 1 & 0 & 0 & 1 & 1 & 0\\
	1 & 0 & 0 & 0 & 1 & 0 & 0\\
	0 & 1 & -1 & 0 & 0 & 1 & -1\\
	1 & 0 & -1 & 0 & 1 & 0 & -1\\
	1 & -1 & 0 & 0 & 1 & -1 & 0
	\end{bmatrix},
	\end{equation} 
	with order $\tilde v_1^0, \tilde v_2^0, \tilde v_3^0, \tilde v_4^0, \tilde
	\tilde v_1^1, \tilde v_2^1,\tilde v_3^1$. It is expected that starting from
	arbitrary proper mixed state (at least in a vertex, \ie{} $\tilde
	\varrho_0=\ketbra{\tilde v_i^j}{\tilde v_i^j}$), we should obtain 
	$\lim_{t\to\infty}\tilde
	\varrho_t = \ketbra{\tilde v_4^0}{\tilde v_4^0}$. Oppositely, one can verify 
	that
	\begin{equation}
	\bar \varrho = \frac{1}{16}\begin{bmatrix}
	5 & 1 & 1 & 0 & -5 & -1 & -1\\
	1 & 1 & 1 & 0 & -1 & -1 & -1\\
	1 & 1 & 1 & 0 & -1 & -1 & -1\\
	0 & 0 & 0 & 2 & 0 & 0 & 0\\
	-5 & -1 & -1 & 0 & 5 & 1 & 1\\
	-1 & -1 & -1 & 0 & 1 & 1 & 1\\
	-1 & -1 & -1 & 0 & 1 & 1 & 1
	\end{bmatrix},
	\end{equation}
	is a stationary state of the evolution. Such state can be obtained starting from
	initial state $\tilde \varrho_0 = \ketbra{\tilde v_1^0}{\tilde v_1^0}$.

	To correct this problem, we propose to add the Hamiltonian $\tilde
	H_{\textrm{rot}}$ which changes the state within the subspace corresponding to
	single vertex, \ie{} if $i\neq j$ then $\bra{\tilde v_i^l}\tilde
	H_{\textrm{rot}}\ket{\tilde v_j^k}= 0$. Now the evolution takes the form
	\begin{equation}
	\frac{\dd}{\dd t}\tilde \varrho =-\ii [\tilde H_{\textrm{rot}},\tilde  
	\varrho] + 
	\tilde L 
	\tilde \varrho 
	\tilde L^\dagger - \frac12 \{\tilde L^\dagger\tilde  L, \tilde \varrho\}.
	\label{eq:GKSL-premature-localization}
	\end{equation}
	We call this Hamiltonian the \emph{locally rotating Hamiltonian}, since it acts
	only locally on the subspaces corresponding to single vertex. We have verified
	numerically that the appropriate Hamiltonian corrects the premature
	localisation. Application of the rotating Hamiltonian in the evolution on the
	graph, presented in Fig. \ref{fig:premature-localization-example-increased},
	gives new stationary state $\ketbra{\tilde v_4^0}{\tilde v_4^0}$. Furthermore,
	numerical analysis show that random non real-valued Hamiltonian corrects the
	undesired effect for an arbitrary random digraph.
	
	In the scope of this paper we will use the Hamiltonian
	\begin{equation}
	\bra{\tilde v_{i}^k}\tilde H_{\textrm{rot}}\ket{\tilde v_j^l} = 
	\begin{cases}
	\ii, & i=j \textrm{ and } l=k+1\mod \indeg (v_i), \\
	-\ii, & i=j \textrm{ and } l=k-1\mod \indeg (v_i), \\
	0, & \textrm{otherwise.}
	\end{cases}\label{eq:example_ham_rot}
	\end{equation}
	
	\subsection{Lack of symmetry on symmetric graphs}\label{sec:symmetry}
	
	\begin{figure}
		\centering 
		\subfigure[\label{fig:nonsymmetry-line}]{
			\includegraphics[width=0.48\textwidth]{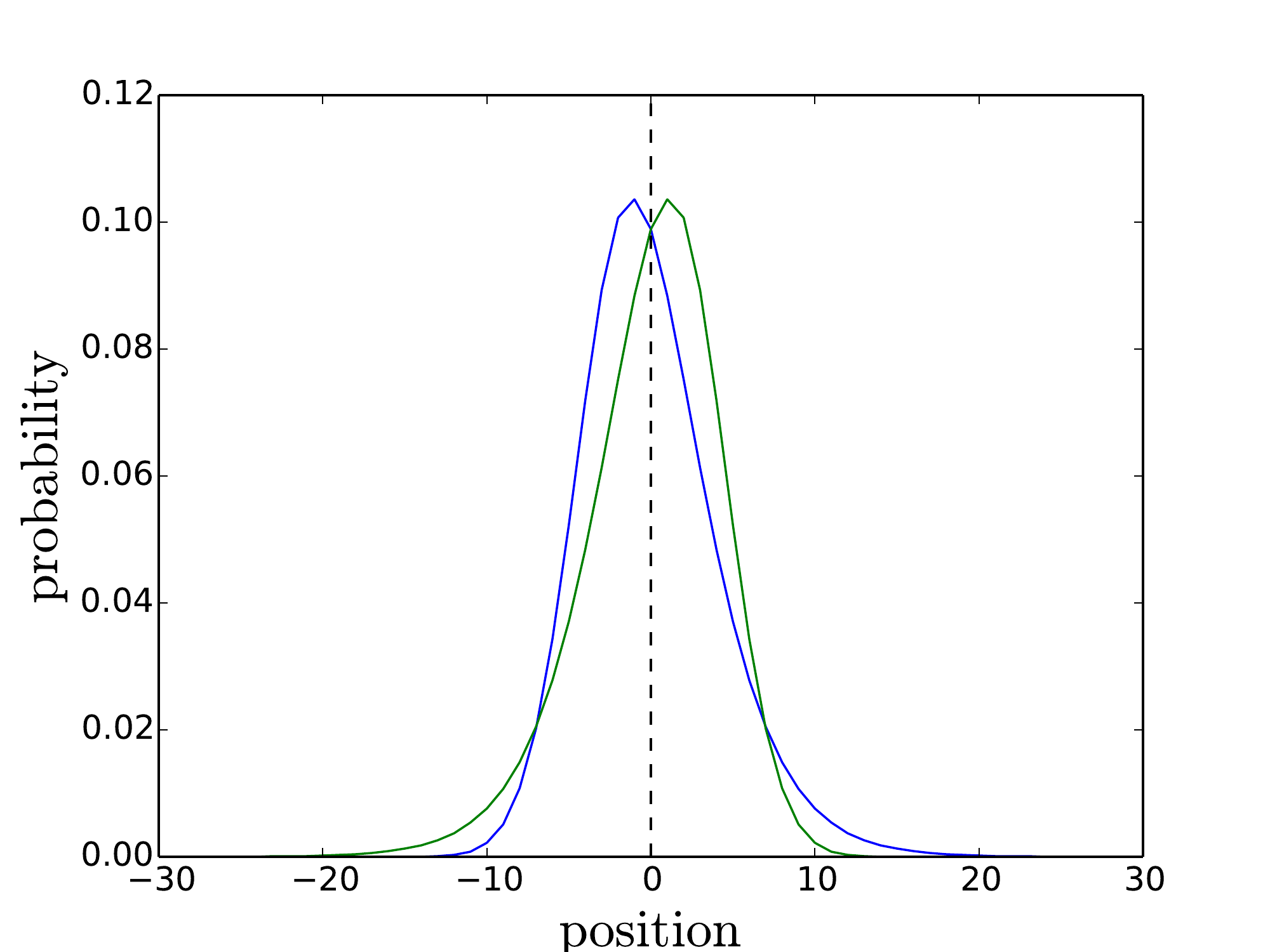}}
		\subfigure[\label{fig:symmetry-line}]{\includegraphics[width=0.48\textwidth]{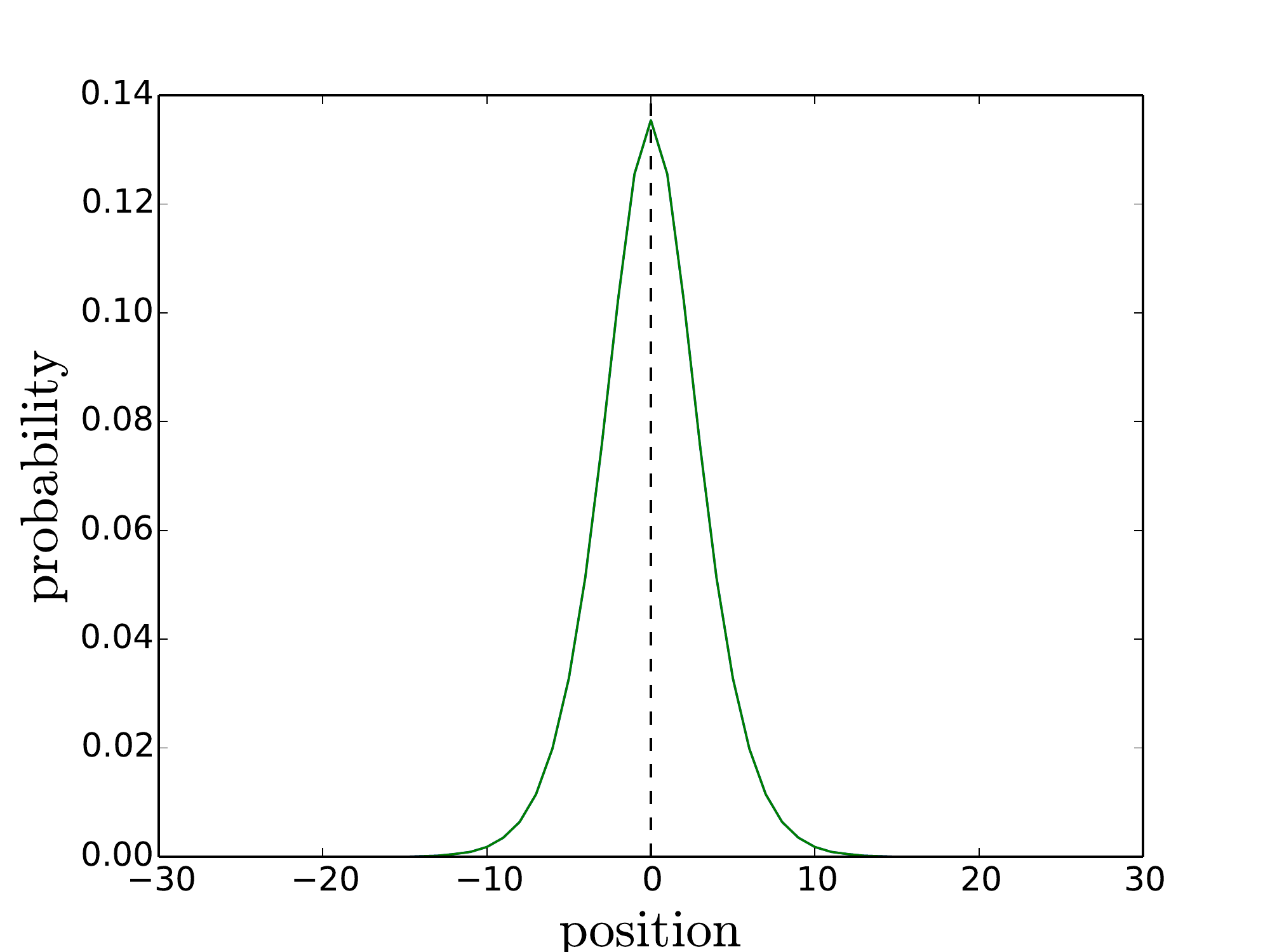}}
		
		\caption{ Probability distribution of the measurement in the standard 
			basis and its reflection by the initial position for $t=20$ on the line 
			segment of length 51, before procedure application Fig. 
			\subref{fig:nonsymmetry-line} and after procedure application Fig. 
			\subref{fig:symmetry-line}. In both cases we start in  
			$\frac{1}{2}(\ketbra{\tilde v_0^0}{\tilde 
				v_0^0}+\ketbra{\tilde v_0^1}{\tilde v_0^1})$. }
	\end{figure}
	
	A further undesired effect  has occurred for some symmetric graphs, where we
	observe the lack of symmetry of the probability distribution. Let us analyse the
	walk described in Eq.~\eqref{eq:GKSL-premature-localization} on undirected
	segment graph. In Fig.~\ref{fig:nonsymmetry-line} we present the probability
	distribution of the position measurement and the reflection of distribution
	according to the initial position. We observe that probability distribution is
	not symmetric with respect to the initial position. 
	
	Removing the locally rotating Hamiltonian doest not removed the asymmetry, hence
	it should come from the Lindbladian part of the evolution. Our analysis of
	transition rates suggests that such behaviour results from non-symmetry of the
	propagation $\tilde \MM_{\tilde v\tilde v}^{\tilde w\tilde z} = \bra{\tilde
		w}\tilde L \ket {\tilde v } \bra{ \tilde v} \tilde L^\dagger \ket{ \tilde z}$. 
	In the case of the Fourier matrix used in our model, values of $\tilde
	\MM_{\tilde v\tilde v}^{\tilde w\tilde z}$ belong to a complex unit circle.
	Since the columns of matrices $A_{i}$, need to form an orthogonal basis, the
	values $\tilde \MM_{\tilde v\tilde v}^{\tilde w\tilde z}$ will be different in
	general. We verified numerically, that there is no single matrix $A$, which can be used for all vertices of enlarged graph $\tilde G$ that removes the asymmetry. Hence we propose a different approach based on adding reflected operator.
	
	Our idea is to add another global interaction Lindblad operator, with
	different $A_i$ matrices which will remove the side-effect. In the case of
	undirected segment, we choose $\tilde \LL=\{\tilde{L}_1,\tilde{L}_2\}$ with
	matrices $A_{i}^{(1)}$ and $A_{i}^{(2)}$ for the $i^{\textrm{th}}$ vertex.
	Henceforth, for $\tilde{L}_1$ we choose for each vertex
	\begin{equation}
	A_i ^{(1)}=\begin{bmatrix}
	1 & 1 \\1 & -1
	\end{bmatrix}
	\end{equation}
	and for $\tilde{L}_2$ we choose for each vertex 
	\begin{equation}
	A_i ^{(2)}=\begin{bmatrix}
	1 & 1 \\ -1 & 1
	\end{bmatrix}.
	\end{equation}
	Finally, the evolution takes the form
	\begin{equation}
	\frac{\textrm{d}}{\textrm{d}t}\tilde \varrho =-\ii [\tilde 
	H_{\textrm{rot}},\tilde  \varrho] + 
	\sum_{\tilde L \in \{\tilde L_1,\tilde L_2\}}\left (\tilde L 
	\tilde \varrho 
	\tilde L^\dagger - \frac12 \{\tilde L^\dagger\tilde  L, \tilde \varrho\} 
	\right ).
	\label{eq:GKSL-symmetric-segment}
	\end{equation}
	Numerical analysis (see Fig. \ref{fig:symmetry-line}) shows, that we obtain
	symmetric probability distribution for arbitrary time $t$. It is possible, that
	for the general graph the choice of different permutations of columns in $A_i$
	matrix as $A_{i}^{(1)}, A_{i}^{(2)}, \ldots$ would suffice.
	
	It is worth noting that for some graphs the symmetrization procedure is not
	necessary. For example, if we evolve quantum stochastic walks on undirected
	perfect $k$-ary tree, with the initial state localized in the root, we can
	observe that the distribution is symmetric due to arbitrary automorphism $A$
	such that $A({\rm root})={\rm root}$. We have not found an explanation for this
	phenomenon yet.
	
	\section{The propagation speed and performance analysis }\label{sec:hurst}
	
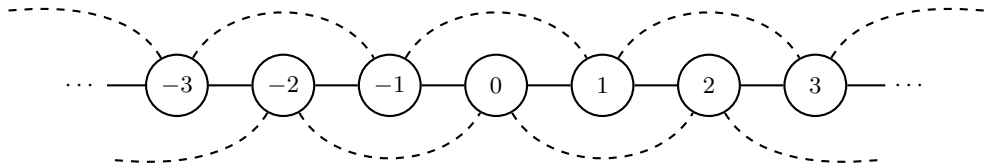
\begin{figure}[h]
		\centering
		
		\begin{tikzpicture}[node distance=1.4cm,thick]
		\footnotesize 
		\tikzset{nodeStyleLine/.style = {circle,draw,minimum size=2.5em}}
		
		\node[nodeStyleLine] (A0)   {$-3$};
		\node[nodeStyleLine] (A1) [right of=A0] {$-2$};
		\node[nodeStyleLine] (A2) [right of=A1] {$-1$};
		\node[nodeStyleLine] (A3) [right of=A2] {$0$};
		\node[nodeStyleLine] (A4) [right of=A3] {$1$};
		\node[nodeStyleLine] (A5) [right of=A4] {$2$};
		\node[nodeStyleLine] (A6) [right of=A5] {$3$};
		\node[] (Eright)   [right=0.5cm of A6] {$\cdots$};
		\node[] (Eleft)   [left=0.5cm of A0] {$\cdots$};
		\draw (A6) -- (Eright) ;
		\draw (A0) -- (Eleft) ;
		
		\node (Am1) [left of=A0] {};
		\node (Am2) [left of=Am1] {};
		\node (Ap1) [right of=A6] {};
		\node (Ap2) [right of=Ap1] {};

		\draw (A0) -- (A1);
		\draw (A1) -- (A2);
		\draw (A2) -- (A3);
		\draw (A3) -- (A4);
		\draw (A4) -- (A5);
		\draw (A5) -- (A6);
		
		\draw[dashed] (A0) to [out=60,in=120,looseness=1] (A2);
		\draw[dashed] (A2) to [out=60,in=120,looseness=1] (A4);
		\draw[dashed] (A4) to [out=60,in=120,looseness=1] (A6);
		\draw[dashed] (A1) to [out=-60,in=-120,looseness=1] (A3);
		\draw[dashed] (A3) to [out=-60,in=-120,looseness=1] (A5);
		
		\draw[dashed,shorten >=1cm] (A1) to [out=-120,in=-60,looseness=1] (Am1);
		\draw[dashed,shorten >=1cm] (A0) to [out=120,in=60,looseness=1] (Am2);
		\draw[dashed,shorten >=1cm] (A6) to [out=60,in=120,looseness=1] (Ap2);
		\draw[dashed,shorten >=1cm] (A5) to [out=-60,in=-120,looseness=1] (Ap1);
		\end{tikzpicture}
		
		\caption{Line graph. Dashed lines correspond to additional amplitude 
			transitions every two nodes coming from the GKSL model.\label{fig:line}}
	\end{figure}

	In \cite{domino2016properties} we have shown that quantum stochastic walk with
	the Hamiltonian operator and the global interaction Lindblad operator yields a
	ballistic propagation. However, because of the spontaneous moralization, the
	graph which was actually analysed was an undirected line with additional edges
	between every two vertices, see Fig.~\ref{fig:line}. Such observation does not
	diminish the results of \cite{domino2016properties}, but gives additional
	incentive to recompute the scaling exponent of stochastic walks on a line by
	using new model. Hence, in this section we reproduce those results using our
	model, to verify if the fast propagation recorded in \cite{domino2016properties}
	for the global interaction case is due to the additional amplitude transitions
	or due to the quantum stochastic walk itself.
	
	To do so we use the scaling exponent $\alpha$. We define it as a scaling
	parameter of the second moment $\mu_2(t)$, \ie{}
	\begin{equation}
	\lim\limits_{t\to\infty}\frac{\mu_2(t)}{C t^{\alpha}} = 1 
	\end{equation}
	for some positive $C$ or equivalently
	\begin{equation}
	\mu_2(t) = 
	\Theta\left(t^{\alpha}\right).
	\end{equation}
	Generally, the greater the value of $\alpha$, the faster the walk propagates for
	large times. In the case of the walk on the line, it is well known that for the
	classical random walk we have $\alpha = 1$ \cite{caldeira1983path}, and for the
	quantum walk $\alpha=2$ \cite{kempe2003quantum}.
	
	In \cite{domino2016properties} we have analysed evolution of the form
	\begin{equation}
	\frac{\dd}{\dd t} \varrho = -\ii (1-\omega)[H,\varrho] + 
	\omega\left (L 
	\varrho 
	L^\dagger - \frac12 \{L^\dagger L, \varrho\} 
	\right ), 
	\end{equation}
	where both $L$ and $H$ are adjacency matrix on the undirected line, and
	$\omega\in[0,1]$ is a free parameter of interaction strength. In
	\cite{domino2016properties} we demonstrated that for $\omega \in [0,1)$ the
	scaling exponent equals two, \ie{} the propagation will follows the ballistic
	regime.
	
	In our case, we consider the model based on the symmetrized quantum stochastic
	walk given by Eq.~\eqref{eq:GKSL-symmetric-segment}. We introduce a Hamiltonian,
	which is an adjacency matrix of the increased graph $\tilde G$, \ie{}
	\begin{equation}
	\bra{\tilde i}\tilde H\ket{\tilde j} = \begin{cases}
	1, & \bra {\tilde i} \tilde L_1 \ket{\tilde j}\neq 0,\\
	0, & \bra {\tilde i} \tilde L_1 \ket{\tilde j}=0,
	\end{cases}
	\end{equation}
	The evolution takes the form
	\begin{equation}
	\begin{split}
	\frac{\textrm{d}}{\textrm{d}t}\tilde \varrho &=-\ii (1-\omega)[\tilde H 
	,\tilde \varrho] +\omega\left 
	(\ii[\tilde 
	H_{\textrm{rot}},\tilde  \varrho] + 
	\sum_{\tilde L \in \{\tilde L_1,\tilde L_2\}}\left (\tilde L 
	\tilde \varrho 
	\tilde L^\dagger - \frac12 \{\tilde L^\dagger\tilde  L, \tilde \varrho\} 
	\right )\right )
	\label{eq:weighted-line-GKSL-model}. 
	\end{split}
	\end{equation}
	
	In \cite{domino2016properties} we examined quantum stochastic walks on an
	undirected line. For the global interaction case we demonstrated that if $\omega
	\in [0,1)$, the scaling exponent equals two, and thus the propagation if
	ballistic. As it was shown in previous sections, in such case, the walk does not
	propagate on the line graph, but on the graph with additional edges between
	every two nodes, see Fig. \ref{fig:line}. Such observation does no diminish the
	results of \cite{domino2016properties}, but gives additional incentive to
	recompute the scaling exponent of stochastic walks on a line, by using our
	corrected model.

\begin{figure}
		\centering
		\includegraphics[width=0.7\textwidth]{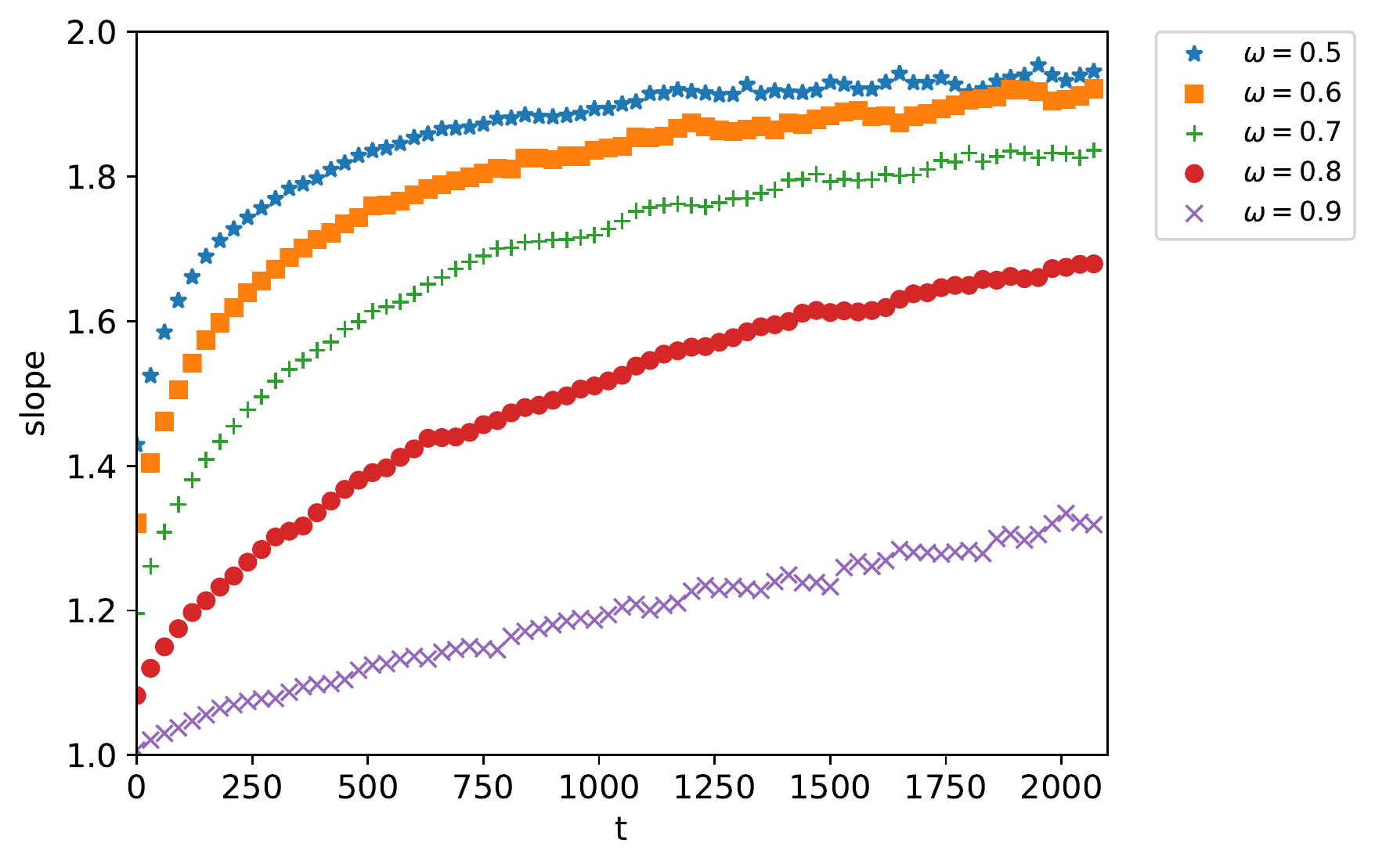}
		
		\caption{Slope of the local regression line for different values of $\omega$,
			respectively from top to bottom: $\omega = 0.5,0.6,0.7,0.8,0.9$. We can see that
			for each value of the $\omega$ slope exceeds $1$ and increases in time.}
		\label{fig:hurst-on-line}
		
	\end{figure}

	To determinate the scaling exponent we use a formula
	\begin{equation}
	\alpha =\lim_{t\to\infty} \frac{\log \mu_2(t)-\log C}{\log t} = 
	\lim_{t\to\infty} \frac{\log \mu_2(t)}{\log t}.
	\end{equation}
	First, we compute the second moment for  $\omega\in\{0.5,0.6,0.7,0.8,0.9\}$ for
	different times from interval $[0,2400]$. Then, we chose 10 consecutive points
	and compute the linear regression of $\log (\mu_2(t))$ vs $\log(t)$. The slope 
	of
	the regression approximates the scaling exponent. The result of numerical
	analysis is shown in Fig.~\ref{fig:hurst-on-line}.
	
	We can see that for each value of $\omega$ the slope increases in time and
	exceeds~$1$, which is the upper bound for classical propagation. The
	result suggests that, at least for some values of $\omega$, we have reproduced
	the ballistic propagation regime $\alpha \lesssim 2$. Moreover one can conclude,
	that the scaling exponent exceeds $1$ even for $\omega$ close to  $1$.
	
	Both results confirm that the fast propagation (ballistic or at least super-diffusive) is the property of global interaction case of the quantum
	stochastic walk and not from the fact that the original model allows additional
	transitions not according to the graph structure. Moreover, it confirms that our
	model preserves the super-diffusive propagation of the walk, at least for some
	interaction strength $\omega$.

	%

	The correction scheme presented in this paper enlarges the Hilbert space used.
	We can bound from above the dimension of constructed space. If the original
	graphs consists of $n$ vertices, with indegree $\textrm{indegree(v)}$ for vertex
	$v$, then the size of enlarged Hilbert space equals $\sum_{v\in
		V}\textrm{indegree}(v)+\#\{v:\textrm{indegree}(v)=0\} $. If
	$\textrm{indegree}(v)=O(f(n))$ for all $v$, the size is roughly $nf(n)=O(n^2)$ in
	the worst case scenario. In the term of number of qubits the additional qubit
	number is $O(\log n)$, hence in our opinion the correction scheme is efficient.
	Comparing to other models \cite{taketani2016physical}, where for each vertex
	there is corresponding qubit, size of our Hilbert space is still small.
	Moreover, while the correction can be applied to arbitrary digraph, real graphs
	are usually sparse, hence the indegree is much more smaller.
	
	Furthermore, our corrections scheme preserves the geometry of the graph, since we
	do not create long range interactions: $\tilde v_i^k$ and $\tilde v_j^k$ are
	connected iff $v_i$ and $v_j$ or $v_i=v_j$. Hence from the geometry point of
	view, our model should not be much harder to realized than original global
	environment interaction case.
	
	In our opinion an interesting further research direction would be generalization
	of our result onto multiplex networks, \ie{} collection of graphs defined on the
	same vertex set. If Lindbladian operator would correspond to different graphs,
	quantum stochastic walks may have interesting new properties comparing to
	classical multiplex random walk. However, in order to define such an evolution,
	one need to generalize the correction scheme in order to remove spontaneous
	moralization on all Lindbladian operators at once.
	
	\section{Conclusions} \label{sec:conclusion} 
	
	In this paper we have shown, that the quantum stochastic walk global interaction
	case allows to transit amplitude not along the graph topology what makes a use 
	of a global interaction case for a ranking algorithms impossible. We have
	discovered, that the graph, on which the evolution actually is made is the moral
	graph, hence we call this effect the spontaneous moralization. Furthermore we
	proposed a new model based on GKSL master equation with global Lindblad operator
	which do not posses such effect. We have recomputed the results from
	\cite{domino2016properties} in context of our model and we have confirmed that
	our corrections do not disturb the ballistic propagation of the walk. This
	demonstrates, the fast propagation corresponds not to classically forbidden
	transitions, but to the quantum stochastic walks model itself.

	\section*{Acknowledgements} Krzysztof Domino acknowledges the support
of the National Science Centre, Poland under project number
2014/15/B/ST6/05204. Adam Glos and Mateusz Ostaszewski were supported by the
Polish Ministry of Science and Higher Education under project number IP 2014
031073. The authors would like to thank Jaros\l{}aw Adam Miszczak for revising
the manuscript.

\bibliographystyle{ieeetr}
\bibliography{stochastic-directed}

\end{document}